\documentclass[preprint2]{aastex}
\usepackage{graphicx}
\usepackage{epsfig}
\usepackage{lscape}
\usepackage{graphicx}
\usepackage{amsmath}
\usepackage[english]{babel}
\usepackage[titletoc,title]{appendix}
\usepackage{txfonts}
\usepackage{times}
\usepackage{natbib}
\begin{document}

\title{Constraining the Warm Dark Matter Particle Mass through Ultra-Deep UV Luminosity Functions at z=2} 
\author{N. Menci$^1$, N.G. Sanchez$^2$, M. Castellano$^1$, A. Grazian$^1$}

\affil{$^1$INAF - Osservatorio Astronomico di Roma, via di Frascati 33, 00040 Monte Porzio Catone, Italy}
\affil{
$^2$Observatoire de Paris, LERMA, CNRS UMR 8112, 61, Observatoire de Paris PSL, Sorbonne Universit\'es, UPMC Univ. Paris 6, 61 Avenue de l'Observatoire, F-75014 Paris, France}

\begin{abstract}

We compute  the mass function of galactic dark matter halos for different values of the Warm Dark Matter (WDM) particle mass $m_X$ and compare it with the  abundance of ultra-faint galaxies derived from the  deepest UV luminosity function available so far at redshift $z\approx 2$. The magnitude limit $M_{UV}=-13$ reached by such observations allows us to probe the WDM  mass functions down to scales close to or smaller than the half-mass mode mass scale $\sim 10^9\,M_{\odot}$. This allowed for an efficient discrimination among predictions for different $m_X$ which turn out to be  independent of the star formation efficiency adopted to associate the observed UV luminosities of galaxies to the corresponding dark matter masses. Adopting a conservative approach  to take into account the existing theoretical uncertainties in the galaxy halo mass function, we derive a robust limit $m_X\geq 1.8$ keV for the mass of thermal relic WDM particles  when comparing with the measured abundance of the faintest  galaxies, while $m_X\geq 1.5$ keV is obtained when we compare with   the Schechter fit to the observed luminosity function. The corresponding lower limit for sterile neutrinos depends on the modeling of the production mechanism; for instance $m_{sterile}\gtrsim 4$ keV holds for the Shi-Fuller mechanism.  We discuss the impact of observational uncertainties on the above bound on $m_X$. As a baseline for comparison with forthcoming observations from the HST Frontier Field, we provide predictions for the abundance  of faint galaxies with $M_{UV}=-13$ for different values of $m_X$ and of the star formation efficiency, valid up to $z\approx 4$.
\end{abstract}

\section{Introduction}

In cosmological models of structure formation, galaxies form from the condensation of baryons within  larger dark matter (DM) halos. These originate from the collapse of  density perturbations in the DM density field, which dominates the matter content of the Universe and drives the collapse and the growth of cosmic structures (see Peebles 1993). In the  paradigm of cold dark matter (CDM), the primordial DM density perturbations have slowly increasing amplitude down to sub-galactic scales, owing to the negligible thermal velocities of the CDM 
particles. This yields an ever-increasing number of collapsed small-mass halos,  resulting into a steep halo mass function $N(M)\sim M^{-2}$ and into cuspy density profiles of DM halos. Although the physics of baryons and the energy injected by Supernovae (Dekel \& Silk 1986) are known to suppress the luminosity-to-mass ratio $L/M$ (see Somerville \& Dav\'e 2014 and references therein) and to expand the DM distribution inside halos (see Mashchenko, Wadsley \& Couchman 2008; Read \& Gilmore 2005; Governato et al. 2012; Ogiya \& Mori 2014; Di Cintio 2014), it is currently matter of debate whether such an effect can simultaneously account for all the observed properties of low-mass galaxies, namely, their flat luminosity (or stellar mass) function $N(L)\sim L^{-1.3}$ up to $z\approx 3$ (see, ee.g., Fontana et al. 2006; Fontanot et al. 2009; Lo Faro et al. 2009; Marchesini et al. 2009; Guo et al. 2011; Parsa et al. 2015), their cored DM density profiles (see, e.g., Alam, Bullock, \& Weinberg 2002; Kuzio de Naray \& Spekkens 2011; de Vega, Salucci \& Sanchez 2014), the low abundance of satellites in Milky Way-like halos (see, e.g., Klypin et al. 1999; Moore et al. 1999; but see also Koposov et al. 2009) and around field galaxies up to $z\sim 1$ (Nierenberg et al. 2013), and the observed 
$L/M$ ratios in the population of dwarf galaxies (see Boylan-Kolchin, Bullock \& Kaplinghat 2012; Moster, Naab, White 2013; Collins et al. 2014; Ferrero et al. 2012; Garrison-Kimmel et al. 2014b; Tollerud, Boylan-Kolchin \& Bullock 2014; Papastergis et al. 2015; Sawala et al. 2015; Pawlowski et al. 2015), while still matching the observed local distribution of rotational velocities which shows only a mild increase $N(v)\sim v^{-1.2}$ at the  low-mass end (Papastergis et al. 2011; Klypin et al. 2014) and the relatively blue colors of satellite galaxies (Kimm et al. 2009; Dave' et al. 2011; Hirschman et al. 2012, Bower et al. 2012, Weinmann et al. 2012; Hirschman et al. 2013). 

In this context, models based on warm dark matter (WDM, see Bode, Ostriker \& Turok 2001), composed by particles with masses in the keV scale (see de Vega \& Sanchez 2010), are receiving increasing attention. In fact, in such a scenario the perturbation power spectrum is suppressed with respect to the CDM case at small scales $\lesssim 1$ Mpc due to the free-streaming of the lighter and faster DM particles.  The corresponding lack of small-scale power results in a reduced dwarf galaxy abundance and shallower inner density profiles naturally matching the observations (Macci\'o et al. 2012; Schneider et al. 2012, Lovell et al. 2012, 2014; Polisensky \& Ricotti et al. 2011; de Vega, Salucci, Sanchez 2014; 
Papastergis et al. 2015; see also Polisenski \& Ricotti 2015 and references therein), while being indistinguishable from CDM on larger scales. Recent works have investigated the impact of  adopting a WDM cosmology on galaxy formation, showing that it can naturally match the observed number of satellites (Nierenberg et al. 2013), the galaxy luminosity and stellar mass distributions over a wide range of masses and redshift (Menci et al. 2012; Benson et al. 2013; Dayal, Mesinger, Pacucci 2015), as well as the star formation properties of low-mass galaxies (Calura, Menci, Gallazzi 2014), thus constituting a viable framework for galaxy formation and even for the evolution of the AGN population (Menci et al. 2013). 

The two most popular classes of WDM particle candidates are thermal relic particles and sterile neutrinos
(for a discussion see Colombi, Dodelson, Widrow 1996; for recent reviews see, e.g., Biermann, De Vega \& Sanchez 2013). While for the former it is possible to derive a one-to-one correspondence with the shape of the power spectrum, for sterile neutrinos the free-streaming length  (and hence the effect on the perturbation spectrum) depends on their production mechanism (see Destri, de Vega, Sanchez 2013a). In fact, these can be produced at the proper dark matter abundance through different processes: i) scattering  due to their mixing with active neutrinos through the Dodelson \& Widrow (1994) mechanism; ii) resonant production including the case of  lepton asymmetry (Shi \& Fuller 1999); iii)  coupling with other fields (Kusenko 2009; Shaposhnikov \& Tkachev 2006). Several astrophysical constraints  
are leading to a rather definite window for the thermal relic mass $1\lesssim m_X/keV\lesssim 4$, yielding a power spectrum corresponding to that produced by (non-thermal) sterile neutrinos with mass $m_{sterile}\approx 4-12$ keV, depending on the production mechanism. Indeed, these constitute the simplest candidates (see, e.g.,  Abazajian 2014) for a dark matter interpretation of the origin of the recent unidentified X-ray line reported in stacked observations of X-ray clusters with the XMM-Newton telescope with both CCD instruments on board the telescope, and of the Perseus cluster with the Chandra observatory (Bulbul et al. 2014; independent indications of a consistent line in XMM-Newton observations of M31 and the Perseus Cluster is reported in Boyarsky et al. 2014).  Such a value is consistent with lower limits $m_X\geq 1$ keV (for thermal relics at 2$\sigma$ level, Pacucci, Mesinger \& Haiman 2013) derived from the density of high-redshift galaxies set by the two objects already detected at $z\approx 10$ by the Cluster Lensing And Supernova survey with Hubble (CLASH).  A similar limit has been derived by Lapi \& Danese (2015) from the observed abundance of 
 $z\approx 8$ galaxies with UV magnitude $M_{UV}\approx -17$, while tighter limits can be inferred only by extrapolating the UV luminosity function to magnitudes as faint as $M_{UV}\approx -13$. 
The thermal relic equivalent mass of sterile neutrinos inferred from X-ray observations is also consistent with the limits set by different authors  ($m_X\geq 2.3$ keV, Polisensky \& Ricotti 2011; $m_X\gtrsim 1.5$ keV,  Lovell et al. 2012, 2015;  Horiuchi et al. 2014) from the abundance of ultra-faint Milky Way satellites measured in the Sloan Digital Sky Survey (see, e.g., Belokurov et al. 2010), but in tension with that derived by comparing small scale structure in the Lyman-α forest of high- resolution ($z > 4$) quasar spectra with hydrodynamical N-body simulations which yields $m_X\gtrsim 3.3$ keV  (Viel et al. 2013). 

At present,  various uncertainties may still affect the constraints (for the points below, see discussions in Abazjian et al. 2011, Watson et al. 2012; Schultz et al. 2014 and references therein; Lovell et al. 2015). Lyman-$\alpha$ is  a challenging tool, and requires disentangling the effects of pressure support and thermal broadening from those caused by the DM spectrum, as well as assumptions on the thermal history of the intergalactic medium and of the ionizing background (see Garzilli \& Boyarsky 2015). 
Additional uncertainties may derive from rendering the  velocity dispersion of the real WDM particles with simulation particles which are  $10^{68}$ times heavier. Also for the comparisons of sub-halos to Milky Way dwarfs, these usually assume a correction factor $\approx  4$ to account for the number of dwarfs being missed by current surveys, and lower correction factors would appreciably weaken the constraints.  Given the potential systematic problems with known astrophysical WDM constraints, and the present lack of experimental determination from particle physics experiments (see, e.g., Lasserre et al. 2014), it is useful to explore different probes. 

A first approach is to compare predicted mass distributions with observed luminosity functions of  galaxies at high redshifts. In fact, the exponential suppression of small-scale structures characteristic of WDM models should yield amplified differences with respect to CDM at early cosmic times, since in any hierarchical model of structure formation low-mass objects are the first to collapse. An observed abundance of galaxies at early epochs can in principle set strong lower limits on the WDM particle mass (e.g. Barkana, Haiman \& Ostriker 2001; Mesinger, Perna \& Haiman 2005). This is the approach taken by Schultz et al. (2014; see also Dayal, Mesinger, Pacucci 2015), who compare 
UV luminosity functions of dropout galaxies at $z>6$ with the predicted mass function of different WDM models, 
after connecting the UV luminosity to the DM halo mass through the abundance matching technique.
 The limit of this approach is constituted by the flux limits of present surveys. Observational abundances at high redshifts $z\approx 6$ can be derived only for relatively bright objects $M_{UV}<-16$, corresponding to DM masses $M\geq 10^{10}\,M_{\odot}$. Thus, at present such an approach does not allow to probe  different WDM models in the low-mass regime  $M\leq 10^9\,M_{\odot}$ where the differences are maximal; 
 as a consequence, relatively loose limits $m_X\gtrsim 1$ keV can be set through such a  method. 
  
In this paper we take a different approach to set stringent limits on $m_X$. We choose to compare with the UV luminosity functions at a lower redshift 
$z\approx 2$ but extending to extremely faint magnitudes $M_{UV}\geq -13$. Although at such redshifts the 
differences among the predicted abundances are smaller than at $z\gtrsim 6$, the gain in magnitude allows to probe the mass distribution down to extremely small DM masses $M_{DM}\leq 10^9\,M_{\odot}$, making possible to obtain tighter and robust limits on the WDM candidate mass. 
In particular, the deep UV imaging of lensing clusters with the {\it Hubble Space Telescope}  (HST) has recently allowed to exploit the large magnifications provided by the cluster potential wells to achieve unprecedented detection limit, detecting galaxies down to $M_{UV}=-13$, i.e., 100 times fainter than previous surveys (Alavi et al. 2014). Although incompleteness correction and cluster mass model can still constitute delicate issues, this strategy of surveying large numbers of background galaxies with deep observations of lensing clusters has already been adopted with deep Hubble imaging of the Frontier Fields beginning in Cycle 21. Thus the approach we propose here will be soon applicable to compare with a much larger amount of  data. As we shall show below, the above detection limits allow to probe the mass function of galaxies predicted by WDM with different particle  masses $m_X$  in the region around and below $M=10^9\,M_{\odot}$ where the different models predict maximally divergent abundances. In addition, as we shall show below, at such galaxy masses the WDM mass functions show a maximum which can be compared with 
the observed abundances to set limits to the particle mass $m_X$ which are in practice independent on the specific baryon physics adopted to connect UV magnitudes to DM masses. 

To compare with observations, we derive the galaxy DM mass function from the Monte Carlo realizations of the history of DM halo and sub-halos (see Menci et al. 2012 and references therein). The processes included in the computation are recalled in Sect. 2, where we also test our model results against recent N-body simulations. In Sect. 3 we connect 
the DM masses to UV luminosity, and parametrise the present uncertainties in the $L_{UV}/M$ relation through a 
star formation efficiency $\eta$. This allows us to compare (Sect. 4) the observed densities derived from the deep UV luminosity function by Alavi et al. (2014) with the DM mass function computed assuming different WDM thermal relic masses ranging from 1 keV to 2 keV, for a wide range of possible values of $\eta$. We devote Sect. 5 to conclusions. 

\section{Method}

\subsection{The History of Dark Matter Halos in CDM and WDM Cosmology}

The backbone of the computation is constituted by the collapse history of DM halos on progressively larger scales. Realizations of such histories are generated through a Monte Carlo procedure on the basis of the merging rates given by the Extended Press \& Schechter (EPS) theory, see Bond et al. (1991); Bower (1991); Lacey \& Cole (1993). In this framework, the  evolution of the DM condensations is determined by 
the  power spectrum $P(k)$ of DM perturbations (in terms of the wave-number $k=2\pi/r$) through the variance of the primordial DM density field. This is a function of the mass scale $M\propto \overline{\rho} r^3$ of the DM density perturbations (and of the background density $\overline{\rho}$) given by 
\begin{equation} 
\sigma^2(M)=\int {dk\,k^2\over 2\,\pi^2}\,P(k)\,W(kr)
\end{equation}
where $W(kr)$ is the window function (see Peebles 1993). For the latter, here we adopt a sharp-$k$ form (a tophat sphere in Fourier space) with a halo mass assigned to the filter scale by the relation $M = 4\pi\,\overline{\rho} (cr)^3/3$ with $c = 2.7$ (Schneider et al. 2013). In fact, both theoretical arguments (Benson et al. 2013) and numerical experiments (see Schneider et al. 2012; 2013; Angulo et al. 2013) show that the resulting mass distributions and EPS merger trees (conditional mass function) provide an excellent fit to N-body results for a wide range of DM masses and redshifts (see Schneider 2015). 

Thus, the linear power spectrum $P(k)$ determines the history of collapse, the evolution, and  the mass distribution of DM halos. For the CDM cosmology we adopt the form $P_{CDM}(k)$ given by Bardeen et al. (1986). The WDM spectrum $P_{WDM}(k)$ is suppressed with respect to the CDM case below a characteristic scale depending on the mass of the WDM particles (and, for non-thermal particles, also on their mode of production; see Kusenko 2009; Destri, de Vega, Sanchez 2013a); in fact, the large thermal velocities of the lighter WDM particles erase the perturbations with size comparable to and below the free- streaming scale $r_{fs}$. In particular, if WDM is composed by relic thermalized particles, the suppression factor can be parametrized as (Bode, Ostriker \& Turok 2001; see also Viel et al. 2005; Destri, de Vega, Sanchez 2013a)
\begin{equation}
{P_{WDM}(k)\over P_{CDM}(k)}=\Big[1+(\alpha\,k)^{2\,\mu}\Big]^{-10/\mu}, 
\end{equation}
$${\rm ~with}~~~~~ 
\alpha=0.049 \,
\Bigg[{\Omega_X\over 0.25}\Bigg]^{0.11}\,
\Bigg[{m_X\over {\rm keV}}\Bigg]^{-1.11}\,
\Bigg[{h\over 0.7}\Bigg]^{1.22}\,h^{-1}\,{\rm Mpc},  \nonumber
$$
where $\Omega_X$ is the WDM density parameter, $h$ is the Hubble constant in units of 100 km/s/Mpc, and $\mu=1.12$; to our purpose the above expression for the suppression is equivalent to that found by Destri, de Vega, Sanchez (2013a). 
A similar expression 
holds for sterile neutrinos provided one substitutes the mass $m_X$ with a mass $m_{sterile}$ adopting proper conversion factors.
Note that different relations between $m_{sterile}$ and $m_X$ hold when different production mechanisms are considered (see Kusenko 2009 for a review, and Lovell 2015 for a recent discussion).
Three typical sterile neutrino models are the  Dodelson Widrow (DW) model, the Shi Fuller (SF) resonant production model (depending on the lepton asymmetry $\mathcal L$) and the neutrino Minimal Standard Model ($\nu$MSM).  The masses of the WDM particles in different production models giving the same power spectrum are related to the thermal relic mass $m_X$ according to the following formulas (Destri, de Vega and Sanchez 2013a): 
$m_{DW} \simeq 2.85$ keV $(m_X/{\rm keV})^{4/3}$,  $m_{SF}\simeq 2.55\,m_X$ (in the case of $\mathcal L=0$), and $m_{\nu\,MSM} \simeq1.9\,m_X$. Such conversions are based on 
the WDM primordial spectrum of fermions decoupling out of thermal equilibrium, computed by solving the evolution Volterra integral equations derived from the linearized Boltzmann-Vlasov equations in De Vega \& Sanchez (2012). Note however that the best-fit expressions for the numerically computed power spectra of non-thermal particles provided by different authors may 
 differ in the literature: for a discussion concerning the accuracy of the above conversion factors and the comparison with previous literature (including Viel et al. 2005) we refer to  Destri, de Vega and Sanchez (2013a).

The characteristic scale of power suppression due to free streaming in WDM can be quantified through the ‘half-mode’ scale at which the WDM transfer function drops to 1/2 (Schneider et al. 2012). This is given by (see Schneider et al. 2012) 
\begin{equation}
M_{1/2}={4\pi\over 3}\,\overline{\rho}\,\Bigg[\pi\alpha\Big(2^{\mu/5}-1\Big)^{-{1\over 2\mu}}\Bigg]^{3}
\end{equation}
where $\overline{\rho}$ is the background density of the Universe. For values of thermal relic masses $m_X=1-2$ keV considered here, one obtains $M_{1/2}=1-5\cdot 10^9\,M_{\odot}$. As shown by several  authors (Schneider et al. 2012, 2013; Angulo et al. 2013), at this scale the DM mass function saturates and starts to turn off. Thus, effectively constraining the WDM particle mass requires probing the abundance of DM halos down to masses $M\approx \,10^9\,M_{\odot}$. 
Below such a mass scale, numerical simulations in WDM cosmology have long been subject to artificial fragmentation (see, e.g., Wang \& White 2007). With the advances in the resolution, it has been shown that 
at scales below $M_{1/2}$ simulations yield a number of 'proto-halos', where the density peak has not fully virialised (Angulo et al. 2013); at these scales, they outnumber the virialized halos. 
In this paper we are not interested in providing a detailed and improved description of the mass function, but rather to derive conservative upper limits to the abundance of low-mass galaxies to compare with observations in order to obtain constraints on the WDM particle mass. Thus,  we consider as an upper limit to the abundance of low-mass galaxies the total (halo+proto-halo) mass function, which is limited from above by the relation  (Angulo et al. 2013): 
\begin{equation}
N=N_{CDM}\,\Bigg(1+{M_{1/2}\over M}\Bigg)^{-1}~.
\end{equation}
Here $N_{CDM}$ is the mass function in the CDM case.
Thus, to derive conservative limits for the WDM particle mass, in the following we shall show the results of our computation for both the virialized halo and the halo+proto-halo population: the first is obtained from EPS computed using the variance in Eq. 1 (with a sharp-k filter for the window function) and the spectrum in eq. 2 (depending on the chosen WDM mass $m_X$); the latter is obtained from CDM merging trees after artificially suppressing the abundance of DM haloes with masses $M\leq M_{1/2}$ with a probability $N/N_{CDM}$ (depending on the chosen WDM mass through the corresponding half-mode mass $M_{1/2}$ in eq. 3).

Since our aim is to compute solid upper limits for the WDM mass function (and hence lower limits for the WDM particle mass), we do not explicitly include the effect of residual thermal velocities (on top of gravitationally induced velocities) derived by Benson et al. (2013). In fact,  
 their implementation is still matter of debate, and several authors argue that their effect is negligible for the models we consider
 (Macci\'o et al. 2012; Shao et al. 2013; Schneider 2013; Angulo et al. 2013). In any case, all authors agree that their inclusion would provide an even larger suppression of the WDM abundances below $M_{1/2}$ (see, e.g., Benson et al. 2013), so that neglecting their effect allow us to derive conservative constraints.

\begin{center}
\vspace{-0.1cm}
\scalebox{0.38}[0.38]{\rotatebox{-0}{\includegraphics{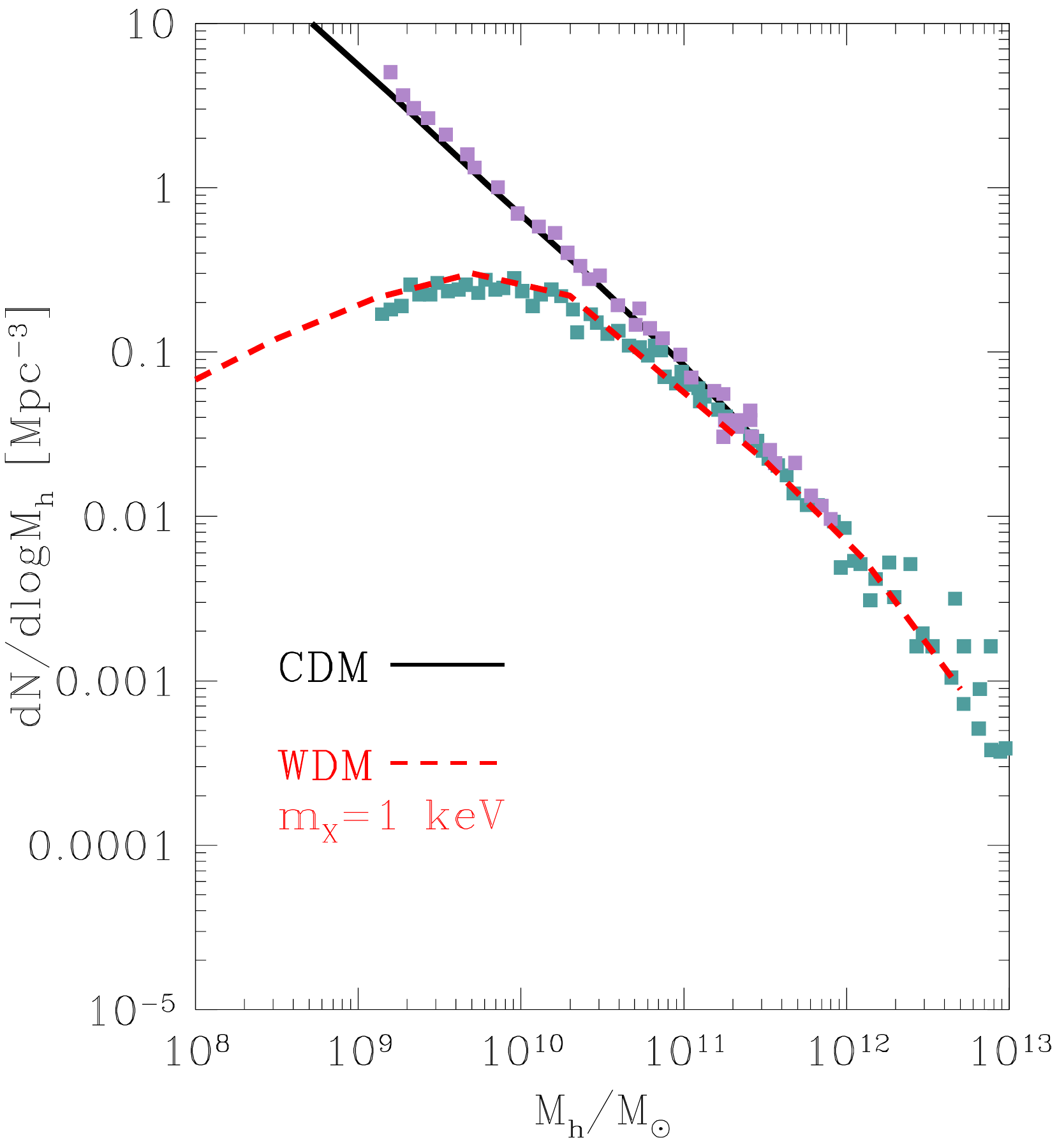}}}
\scalebox{0.36}[0.36]{\rotatebox{-0}{\includegraphics{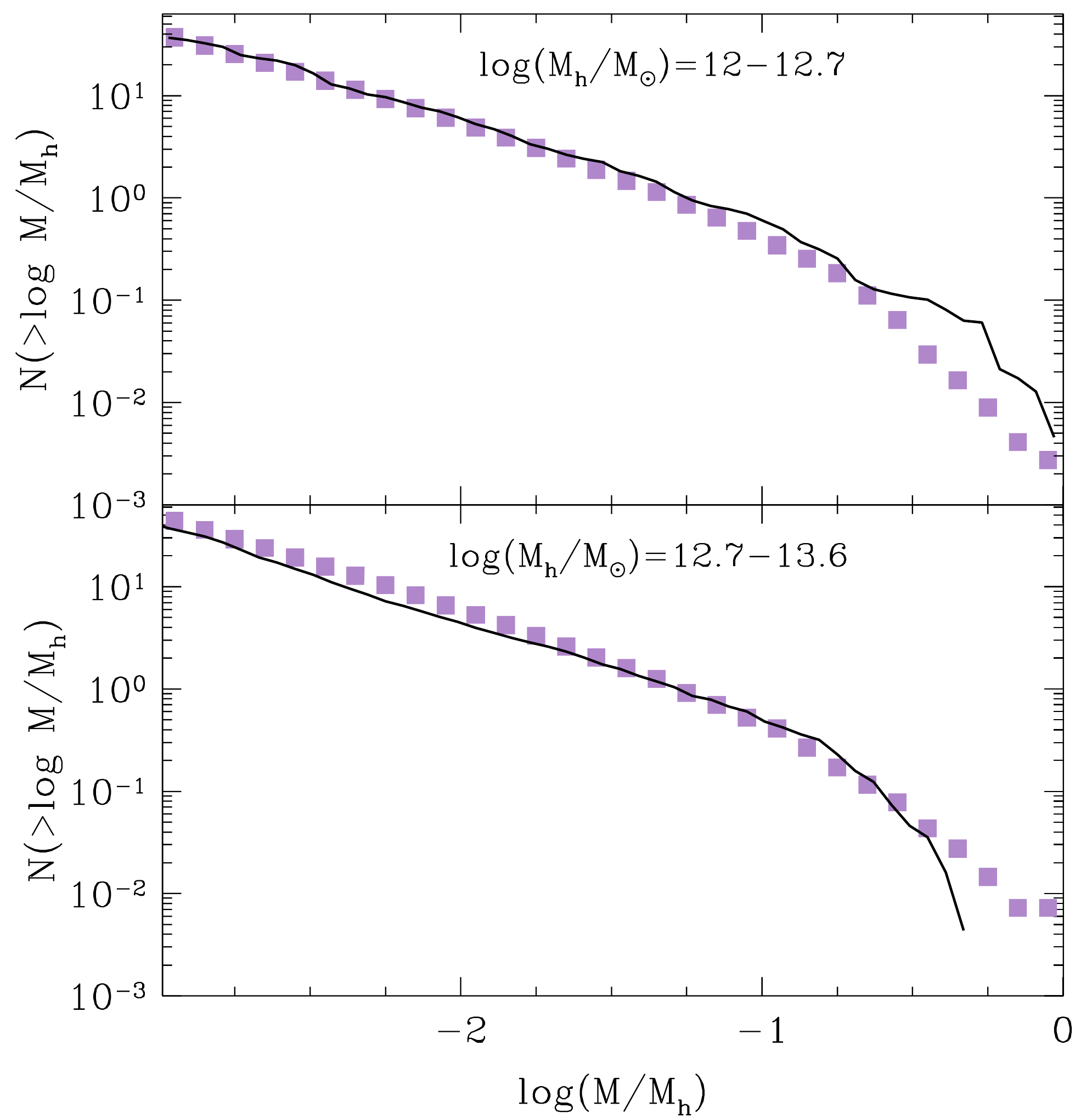}}}
\end{center}
\vspace{-0.3cm }
 {\footnotesize 
Fig. 1. To illustrate the reliability of our model in predicting the galactic DM mass functions we compare with 
different existing simulations. Top Panel: The  mass functions of isolated halos at $z=2.4$ computed from our model for the CDM case (solid black line) and for WDM with $m_X=1$ keV (dashed line) assuming a sharp-k filter. We compare with available simulations at the same redshift by Schneider et al. (2013) for the same DM models. Bottom Panels: To test our computation of the dynamical evolution of sub-halos, we compare with existing 
simulations in the CDM case. The cumulative mass function of sub-halos computed in our model in the CDM case (solid lines) are compared with results from the CDM Bolshoi simulations (Klypin et al. 2011) for the two different ranges of host halo masses, as indicated in the panels. 
\vspace{0.2cm}}

After computing the histories of host DM halos with the EPS approach described above, we follow the 
history of collapsed DM clumps once they are included in larger virialized host DM halos (details on the computation are given in Menci et al. 2012; 2014). Such sub-halos, associated to the single galaxies,  are tracked in our Monte Carlo simulations following the  canonical procedure adopted by semi-analytic models of galaxy formation: they may survive as satellites, or merge to form larger galaxies due to binary aggregations, or coalesce into the central dominant galaxy due to dynamical friction  (see, e.g., Gan 2010; Sommerville \& Dav\'e 2014 and references therein).

To test our computation we first compare the resulting mass function of isolated halos (host halos) against existing N-body simulations at $z=2.4$, close to the redshift range we shall focus on when we compare with  observed luminosity functions. As shown in the left panel of Fig. 1, the agreement is excellent for isolated virialized halos using a sharp-k filter in the computation of the variance in eq. 1. Testing our computation for the dynamical evolution of sub-halos associated to individual galaxies also leads to a very good agreement with the results from the Bolshoi simulations (Klypin et al. 2011) at the faint end of the distributions for different host halo masses (right panels in Fig. 1); an analogous comparison with the sub-halo mass function resulting from the Millenium simulations is presented in Nieremberg et al. (2013), and also shows a good matching.

Thus, our computation provides reliable, tested results that can be used to predict galaxy mass functions for 
different WDM models, and to compare with the observed UV luminosity function at $z=2$. To this aim, we need to quantify the relation between the UV magnitude of galaxies and the mass of their DM halo, as we discuss in the next section.

\subsection{Relating UV luminosities to DM masses}

The deepest LF derived at $z=2$ so far are those by Alavi et al. (2014) extending down to $M_{UV}=-13$ at $z\approx 2$. 
The large density (per unit magnitude) $N\approx 2$ Mpc$^{-3}$ of galaxies at such faint magnitude constrains  
 the abundance of corresponding DM halos predicted by different DM models. To estimate the range of  masses of the DM halo hosting such low-luminosity galaxies we start from the well established relation between UV magnitude and the star formation rate (expressed in $M_{\odot}\,yr^{-1}$, see Madau et al. 1998)
\begin{equation}
log\,\dot m_*=log\,\Big(m_*/\tau_*\Big)=-0.4\,\Big(M_{UV}+18.16-9.97\,E_{B-V}\Big)
\end{equation}
where $E_{B-V}$ is the color excess (assuming a Calzetti extinction law), and the first equality follows from expressing the star formation rate $\dot  m_*=m_g/\tau_*$ in terms of the gas mass $m_g$ and of the gas conversion timescale $\tau_*$ (Schmidt 1959; Kennicutt 1998; see Santini et al 2014 and references therein). 
To connect the above expression to the mass of the galactic DM halos we define the star formation efficiency factor 
\begin{equation}
\eta=(m_*/M)\,(m_g/m_*)/(10^8{\rm yrs}/\tau_*)
\end{equation}
Substituted in the previous equation, this allows to recast the Eq. 4 in the form 
\begin{equation}
log\,M/M_{\odot}=-0.4\,\Big(M_{UV}+18.16\Big)+8-log\,\eta
\end{equation}
where we have taken  $E_{B-V}\approx 0$ (as estimated by Alavi et al. 2014 for the faintest galaxies in their sample). Different observations allow to estimate the range of possible values for $\eta$. 

Measurements of the conversion timescale at $z=2$ yield 
$\tau_*=0.5-2\cdot 10^8$ yrs (Daddi et al. 2010; Genzel et al. 2010; Santini et al 2014; Silverman et al. 2015). 

The gas-to-stars ratio $m_g/m_*$ is usually expressed in terms of the  total gas fraction $f_g=m_g/(m_g+m_*)$; recent estimates indicate that such a fraction increases with redshift and with decreasing stellar mass; measurements derived from CO surveys  (see Tacconi et al. 2013; Conselice et al. 2013;  Silverman et al. 2015) and from deep IR {\it Herschel} observations of dust emission (Magdis et al. 2012; Santini et al. 2014) yield a $f_g=0.2-0.65$ for the faintest galaxies at $z=2$, corresponding to a range $m_g/m_*=f_g/(1-f_g)=0.4-2$. 

As for the star-to-DM ratio $(m_*/M)$, this is subject to large uncertainties, as shown by several existing papers. Works based on the extrapolation of abundance matching relations between the observed stellar mass function and the CDM halo mass distribution (see Moster et al. 2013; Behroozi et al. 2014) show that at $z=2$ it takes values in the range $-3\leq log(m_*/M)\leq -2$, a range supported also by high-resolution hydrodynamical N-body simulations (see Hopkins et al. 2014) in the CDM framework. However, larger values are indicated by the kinematics of dwarf galaxies (Sawala et al. 2011; Ferrero et al. 2012) and in particular of the Milky Way and M31 dwarf spheroidal galaxies (see Sawala et al. 2015 and references therein; Di Cintio et al. 2015; Papastergis et al. 2015). 
We note in WDM cosmology the $m_*/M$ ratio takes values systematically larger than in the CDM case, both when  this is estimated through the abundance matching technique (due to the flatter shape of the mass function of DM halos in WDM cosmologies) and from estimates based on the galaxy kinematics (due to the lower concentration of WDM halos; de Vega, Salucci, Sanchez 2014; Papastergis et al 2015). Here we shall conservatively consider the whole interval $-3.2\leq log(m_*/M)\leq -1.5$ as compatible with present uncertainties for all  the DM models presented here. 

Given the above observational constraints we obtain the interval $-3.6\leq log\,\eta\leq -1$
for the allowed range of the combination $\eta$ at $z=2$. This corresponds to a range $10^7\leq M/M_{\odot}\leq 3\cdot 10^9$ for the possible values of the halo mass of galaxies associated to $M_{UV}=-13$ at $z=2$, and - according to the discussion above - 
to stellar masses $10^5\lesssim m_*/M_{\odot}\lesssim 10^8$. 

We underline that the above measurements for $\tau_*$, $f_g$, $m_*/M$ are obtained for significantly brighter galaxies than $M_{UV}=-13$. 
We note, however, that similar extrapolations based on $\dot m_*-m_*$ relation at $z\approx 2$ (the "main sequence", see Noeske et al. 2007;  Speagle et al. 2014)
yield consistent results. Measurements of such a relation from recent large  surveys 
yield $log\,m_*/M_{\odot}=7.1\pm 0.7$ for galaxies with star formation rates corresponding to $M_{UV}=-13$; direct fitting the $M_{UV}$-$m_*$ relation for objects in the $1.5<z<2.5$ range in CANDELS
GOODS-South (on the basis of the available release of photometric redshifts and
rest-frame properties from Dahlen et al. 2013 and Santini et al. 2015) yields $log\,m_*/M_{\odot}=6.7\pm 0.5$ for galaxies with $M_{UV}=-13$ (notice that the uncertainties quoted above do not include the possible effects of bursty star formation, which could enlarge the scatter in the star formation main sequence for low stellar masses, 
see Dominguez et al. 2015).

Thus, we are confident that the range $10^7\leq M/M_{\odot}\leq 5\cdot 10^9$ covers the possible values of DM mass of galaxies corresponding to $M_{UV}=-13$. Comparing the observed abundance of such galaxies at $z=2$ with predictions for the abundance of the corresponding DM masses in different DM models constitutes the subject of the next section. 

\section{Results}

To effectively discriminate among different WDM models we need to probe the abundance of galaxies 
down to DM masses $M\sim 10^7-10^9$ $M_{\odot}$, corresponding to UV magnitudes as faint as 
$M_{UV}\geq -14$. This is not trivial: even with 
the new Wide-Field Camera 3 (WFC3) on the HST selecting star-forming galaxies  via the Lyman break technique at $1 < z < 3$ (Oesch et al. 2010; Hathi et al. 2010; Parsa et al. 2015) the detection is limited by the depth of the shallow UV imaging to galaxies with absolute UV magnitude (measured at 1500 ${\AA}$) brighter than $M_{UV}=−19$. A recent burst in the study of ultra-faint galaxies has been made possible 
by using foreground massive systems as lenses to magnify background galaxies. This strong gravitational lensing preserves  surface brightness while spreading out  the emitted light over a larger area and magnifying it. Over the last decade, this has been used to study individual lensed galaxies in great detail (e.g., Pettini et al. 2002; Siana et al. 2008, 2009; Stark et al. 2008; Jones et al. 2010; Yuan et al. 2013). This approach has allowed
 to survey faint star-forming galaxies at $z\approx 2$ behind the massive cluster A1689, reaching magnitudes 100 times fainter than previous surveys at the same redshift (Alavi et al. 2014). Since  the mass distribution of A1689 is well constrained, it was possible  to calculate the intrinsic sensitivity of the observations as a function of source plane position, allowing for accurate determinations of the effective volume as a function of luminosity, and hence for a determination of the luminosity function down to $M_{UV}=-13$. 
 
To compare with the above measurements, we compute the mass function of galactic DM halos at $z=2$ as described in Sect. 2.1 for the CDM case and for WDM models with four different masses for the candidate DM particle (in terms of the  thermal relic mass): $m_X=1$ keV, $m_X=1.5$ keV; $m_X=1.8$ keV; $m_X=2$ keV; in all cases we adopted a matter density parameter $\Omega_0=0.3$, a baryon density parameter $\Omega_b=0.05$, a Dark Energy density parameter $\Omega_{\Lambda}=0.7$, and $h=0.7$. 
The comparison with the observed abundance of faint galaxies with $M_{UV}=-13$ is performed considering different values of the star formation parameter in the range $-3.6\leq log\,\eta\leq -1$ for converting observed UV magnitudes (at 1500 ${\AA}$) to DM masses after eq. 2. The results are shown in Fig. 2, where we also illustrate the effects of the theoretical uncertainties (sect. 2.1) in the computation of the DM mass function; these are represented by the solid areas whose lower and upper bounds correspond to assuming a sharp-k filter or the suppression factor $N/N_{CDM}$.

\begin{center}
\vspace{-0.cm}
\hspace{-0.2cm}
\scalebox{0.38}[0.38]{\rotatebox{-0}{\includegraphics{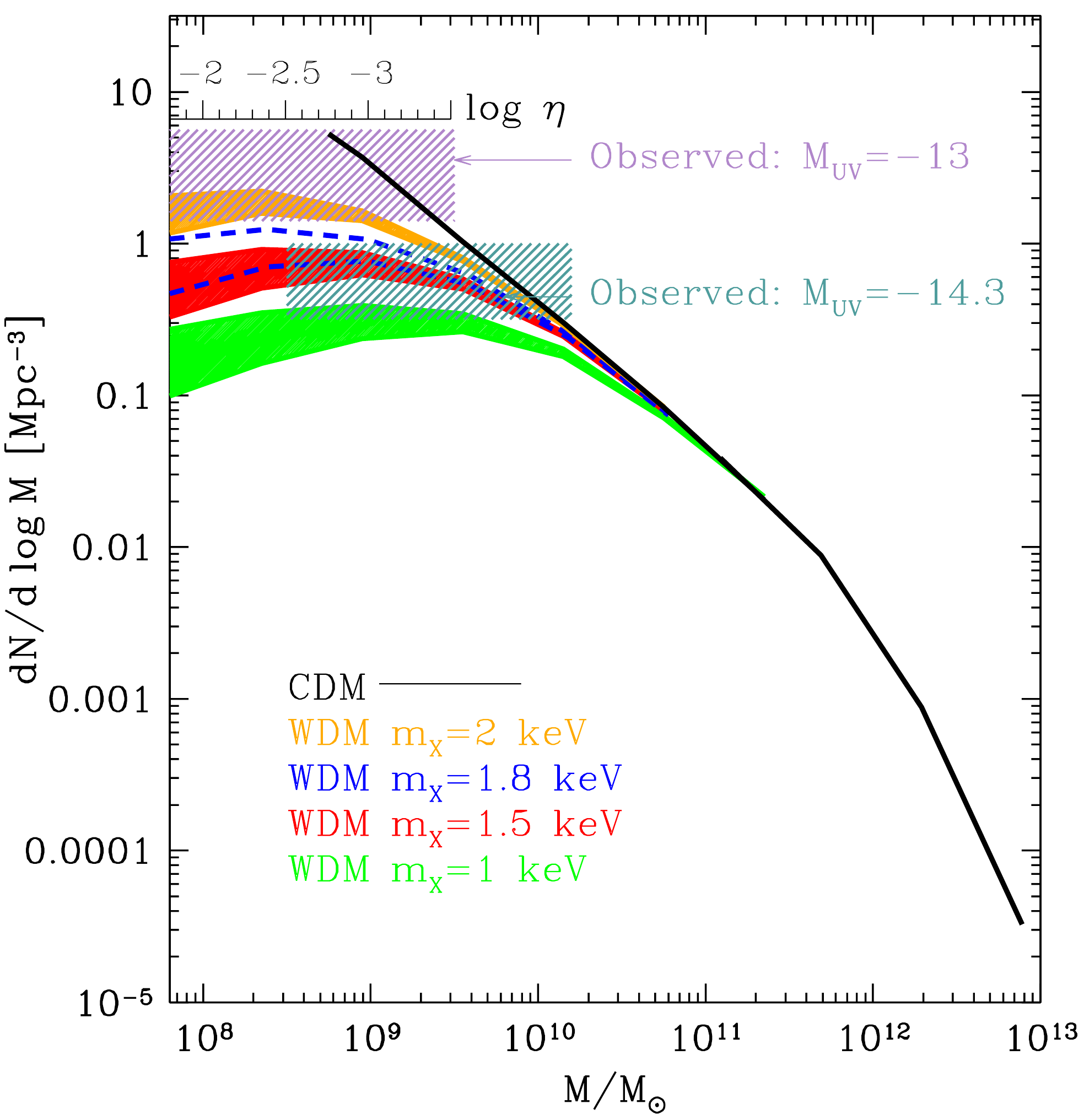}}}
\end{center}
\vspace{-0.4cm }
 {\footnotesize 
Fig. 2. The predicted mass functions of galactic DM halos at $z=2$ corresponding to different thermal relic masses $m_X$ are compared with the observed abundance of faint  galaxies from Alavi et al. (2014) with $M_{UV}=-13$ (upper 
hatched rectangle) and $M_{UV}=-14$ (lower hatched rectangle). The latter have been converted from number density per unit magnitude to 
 number density per unit $log M$. The height of the rectangles corresponds to the Poissonian errorbar given in Alavi et al. (2014), while their horizontal size corresponds to different values of the star formation parameter $\eta$ (see text), as indicated by the small horizontal axis. 
For each value of $m_X$, we bracket the uncertainties in the computation of the DM mass functions (discussed in Sect. 2.1) through the lower and upper bounds of the shaded regions, which correspond to assuming a sharp-k filter or the suppression factor $N/N_{CDM}$ (accounting also for proto-halos). 
\vspace{0.2cm}}

The comparison (even allowing for an additional uncertainty of $\sigma_{CV} \approx 0.13$ in the observed density due to cosmic variance, see Alavi et al. 2014) sets a firm lower limit $m_X\geq 1.8$ keV for the value of the thermal relic mass: the corresponding value of the sterile neutrino masses depends on the production scenarios (as discussed in Sect. 2.1) $m_{sterile}\gtrsim 4.6$ keV for resonantly produced sterile neutrinos. WDM models with smaller values of the particle mass predict too low abundances compared to the observed value 
measured by Alavi et al. (2014) for any choice of the star formation efficiency $\eta$.  On the other hand, particle masses $m_X\sim 2-3$ keV yield galaxy number densities consistent with observations for a wide range of $\eta$, while in the limit of $m_X\gg 1$ we recover the generic CDM result that very inefficient star formation is required to match the observed galaxy abundances.
We stress that our results are {\it conservative}  and are robust with respect to present uncertainties in the modelling of the WDM mass distribution and - most 
important - with respect to the details in the baryonic physics. In particular, they are robust with respect to: 
\newline
i) The specific value of the star formation efficiency $\eta$ and consequently the effect of baryonic processes 
(like star formation, feedback) determining the efficiency of star formation for given mass of the host DM halo. 
This is due to the fact that the ultra-deep observations we are comparing with allow us to probe the DM halo mass 
function in the mass range around the half-mode mass where the DM mass functions  are characterized by a maximum value.  \newline
ii) The modeling of the effects of residual DM dispersion velocities on the mass function of DM haloes (see sect. 2.1): while in our computation these have not been included, their effect would be to provide a sharper decrease of the mass function at small masses (see, e.g., Benson et al. 2013), thus yielding tighter constraints.
\newline
iii) The kind of DM clumps hosting the UV emitting galaxies. In fact,  the upper boundaries of the solid filled  regions in Fig. 2 correspond to predictions including also proto-halos (see discussion in Sect. 2.1). \newline
iv) The possible effects of UV background and reionization. Indeed, such effects go in the direction of  further suppressing the abundance of galaxies in low-mass halos (Sawala et al. 2015), so our  limits are conservative with respect to these processes. 

We also note the importance of having pushed the magnitude limit of the measured luminosity functions to faint values.
Indeed, limiting the computation to $M_{UV}\leq -14-15$ (see Fig. 2) would only allow to obtain $m_X\geq 1$, as indeed already obtained by Smith et al. (2014) based on the UV luminosity functions at $z>4$. 

While our results are robust with respect to uncertainties in the modelling of baryon physics and of the halo mass distribution, they descend from the observed number densities derived by Alavi et al. (2014) for the faintest galaxies in their sample. Thus, a delicate issue is constituted by the analysis they adopted to measure the luminosity function at the faint end. 

Uncertainties on the result of Alavi et al. (2014) can
be due to the selection criterion of the $z=2$ galaxies, to completeness corrections,  and to the estimates of the magnification factors. 
As for the first, the Alavi et al. (2014) galaxies have been selected through the widely used
UV-dropout method; further check with spectroscopic redshifts
of a sample of (intrinsically faint) magnified sources
demonstrated that this color criterion is able to recover 75\% of the
galaxies with spectroscopic redshifts $1.8<z_{spec}<2.4$, thus demonstrating the robustness of
the adopted method.
As for the correction for incompleteness at faint magnitudes, this has been derived
 through detailed simulations: Alavi et al. (2014) studied the completeness as
a function of redshift, magnitude and magnification of the sources by
a Monte Carlo approach, randomly varying also the IGM transmission and
the dust attenuations. In addition, they take into account the photometric
uncertainties, the size distribution of the galaxies and the
instrumental effects (e.g. charge transfer inefficiency). At $z\sim 2$
the typical value for the completeness of galaxies of $M_{UV}\sim -14$
with a magnification of 5 magnitudes is $\sim 40\%$. Under these
considerations, we can conclude that the incompleteness correction 
is robust even at their faintest limits.
The uncertainties on the magnification can be the dominant source of
errors in Alavi et al. (2014) results. Indeed, one galaxy at $M_{UV}\sim -14$ 
in the sample has a
magnification of 8 magnitudes (a factor $>1000$ in luminosity). While, the
reliability of this galaxy is not discussed in detail in Alavi et al. (2014), 
the luminosity function at $M_{UV}=-14$ and $M_{UV}=-13$ we compare with 
is derived from 4 galaxies in total (2 for each mag bin) with much lower magnifications, 
between 30 and 300 in luminosity.  
Thus, the uncertainties in the observed number densities may be larger than those 
related to the Poissonian noise: in  fact, the Schecher fit to the whole luminosity function 
in Alavi et al. (2014), when computed in the faintest magnitude bin $M_{UV}\approx -13$, 
is centered significantly lower than the measured number density.

To discuss the impact of  the uncertainties in the measurements by Alavi et al. (2014) and to the aim of providing a baseline for future observations,  we compare in Fig. 3  the measured density of galaxies with $M_{UV}=-13$ with that  
expected for different WDM particle masses (expressed as above in terms of the equivalent thermal relic mass $m_X$); the color code shows the star formation efficiency parameter $\eta$ adopted to convert from the DM mass function to the luminosity function. To discuss the impact of uncertainties in the measurement by Alavi et al. (2014), we show both the 
  Poissonian errorbars (hatched region) and (as a dashed line) the number density corresponding to the Schechter fit of the whole sample in  Alavi et al. (2014) computed at $M_{UV}=-13$. Relying on the number density corresponding to the binned data in Alavi et al. (2014) implies a  limit $m_X\geq 1.8$ keV for the mass of the thermal relic WDM particle,  while assuming a larger uncertainty covering the whole range between the binned data and the Schechter fit yields a weaker but still significant limit $m_X\geq 1.5$ keV.  
 
\begin{center}
\vspace{0.cm}
\scalebox{0.36}[0.36]{\rotatebox{-0}{\includegraphics{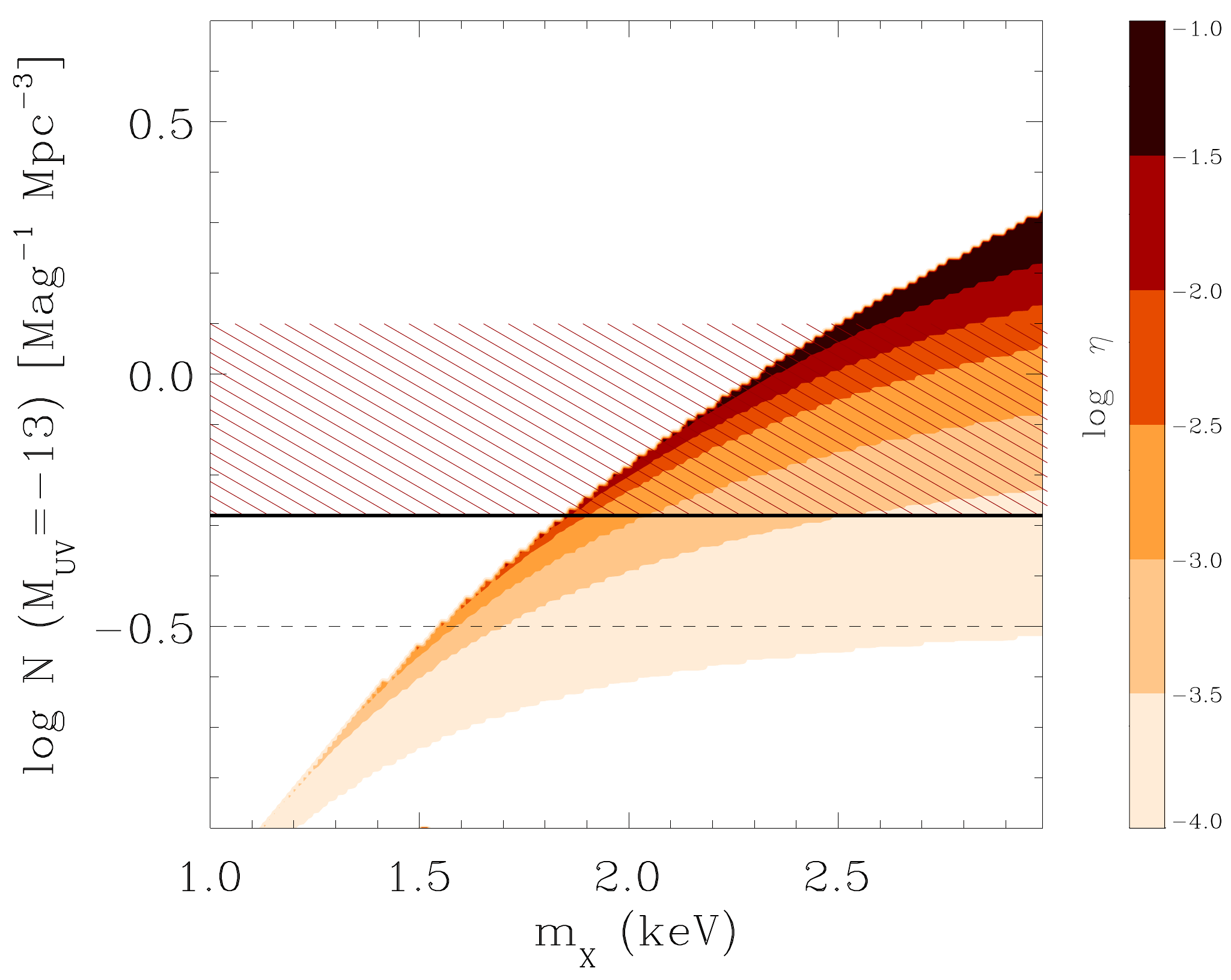}}}
\end{center}
\vspace{-0.2cm }
 {\footnotesize 
Fig. 3. The predicted number density of galaxies per unit magnitude at $M_{UV}=-13$ at $z=2$ is shown as a function of different WDM particle mass (equivalent thermal relics mass $m_X$). The color code shows the star formation efficiency parameter $\eta$ adopted to convert from DM mass function to the luminosity function. The number density corresponding (within their $1\sigma$ errorbars) to the 
measurement by Alavi et al. (2014) is shown as a hatched region. The dashed line shows the number density corresponding to the Schechter fit of the whole sample in  Alavi et al. (2014) computed at $M_{UV}=-13$.
\vspace{0.2cm}} 
 
 Next results from Frontier Field surveys will allow for a confirmation, or even for a strenghtning, of the limits on $m_X$. 
 In fact, although the lensing cluster 
Abell 1689 is at the moment the most studied
cluster with deep HST images and with the most constrained
magnification map (43 confirmed multiple systems with 24 measured 
spectroscopic redshifts, see Limousin et al. 2007; Coe et al. 2010), this database will be probably superseded
soon  by the advent of the HST Frontier Field initiative. In this respect, we stress that, although in this paper we focus on the comparison of different DM models with the measurements by Alavi et al. (2014),  the strategy of surveying large numbers of background galaxies with deep observations of lensing clusters has been adopted with deep Hubble imaging of the Frontier Fields. Thus, the results presented here will constitute a reference baseline for deriving limits on WDM particles from such surveys in the near future.
 
 For the moment, we note that matching a  given observed density 
requires decreasing values of the $\eta$ parameter for increasing values $m_X$. In particular, we recover the 
well-known result that the  CDM  limit corresponding to large $m_X\gg 1$ keV is associated to very inefficient  
star formation efficiencies $log\,\eta\lesssim-3$ (unless a severe suppression of galaxy formation takes place in low mass DM halos with $M<10^9\,M_{\odot}$, see Sawala et al. 2015). Note that although the results in the plot have  been computed at $z=2$ they remain practically valid up to $z=4$, since the low-mass end of the mass function remains almost unchanged in such a range of redshifts.
 
We stress that the masses $m_X$ in Fig. 3 are the equivalent thermal relic masses. We note in particular that  a sterile neutrino with a mass $m_{sterile}=7$ keV,  constituting the best  candidate for a DM interpretation of the origin of the recent unidentified X-ray line reported in stacked observations of X-ray clusters (Bulbul et al. 2014; Boyarsky et al. 2014), corresponds (in the case of Shi-Fuller sterile neutrinos) to a thermal relic mass $m_X\approx 2.8$ keV, which is consistent with the observed abundance of galaxies at $M_{UV}=-13$ for a wide range of star formation efficiency $-2.5\leq log\,\eta\leq -1.5$. Notice that -  
since for $m_X\gtrsim 4$ keV the WDM scenario is nearly indistinguishable from Cold Dark Matter on the scales relevant for galaxy formation - our results define an effective window $1.8\leq m_X\lesssim 4$ keV for the (thermal relic) mass of WDM particles. Finally, the above limits are in agreement with the lower bound obtained from applying the Thomas-Fermi theory to compact dwarf galaxies (Destri, de Vega, Sanchez 2013b).

\section{Conclusions}

We showed how the magnitude limits $M_{UV}=-13$ of present and forthcoming UV luminosity functions allow to probe the number density of DM halos close to - or even below - the half-mode mass of Warm Dark Matter scenario, corresponding to scales $M\sim 10^9\,M_{\odot}$. This provides strong constraints on the mass of the WDM particles which are in practice unaffected by  the details of the baryon physics determining the $L/M$ ratio which at present is measured only for brighter galaxies. In fact,  as long as the number density of ultra-faint galaxies is high
enough, our approach yields constraints on the WDM particle mass that are independent on the star formation efficiency $\eta$.

We applied our method to the  observations by Alavi et al. (2014), who use a foreground galaxy cluster as a lens to magnify background galaxies, so as to measure UV luminosity functions down to the above 
magnitude limits. The measured abundance of galaxies in the faintest magnitude bin $M_{UV}\approx -13$ 
yields a lower limit $m_X\geq 1.8$ keV for the mass of WDM thermal relic particles, while a limit $m_X\geq 1.5$ keV is obtained when we compare with 
 the Schechter fit to the observed luminosity function presented in Alavi et al. (2014).

   The corresponding lower limit for the sterile neutrino mass depends on the production model; for instance, 
 the limit  $m_X\geq 1.8$ keV  for thermal relics corresponds to  $m_{sterile}\gtrsim 4.6$ keV in the case of  the Shi Fuller model with vanishing lepton asymmetry (see the discussion in sect. 2.1 for the conversion factors when other models are considered). The above limit is conservative and robust with respect to the theoretical uncertainties still affecting the computation of the mass function of galactic DM clumps (effects of residual thermal velocities, inclusion of proto-halos) and with respect to the details of the baryon physics (star formation efficiency $\eta$ in a given DM halo, see sect. 2.2). 
Tighter constraints will be achievable when observational determinations of $\eta$ at $z=2$ will be available for galaxies with $M_{UV}=-13$ (see Fig.  3), or when  the number density of such faint galaxies at higher redshift $z\approx 5-8$ will be measured or extrapolated: the latter approach has been proposed by Lapi \& Danese (2015),  who suggest to constrain the abundance of such faint galaxies at high redshift by requiring 
them to satisfy the Planck constraints on reionization.

 Of course, the above limit $m_X\geq1.8$ keV relies on the statistics of ultra-faint field galaxies observed through the magnification provided by lensing clusters (see discussion in sect 3), which at present is limited to A1689.  
However, the present statistical limitations will be  superseded  by the completion of the HST Frontier Field initiative. The Frontier Field campaign (e.g. Coe et al. 2015) is
collecting observations of 6 strong lensing clusters with 7 HST optical and
near infra-red  bands to a 5$\sigma$ depth of ~28.5 (AB magnitudes). Most importantly, large efforts
are ongoing to complement HST observations with ancillary imaging data
from space-based UV (Siana et al. 2014, similarly to the Abell-1689 data
discussed in Alavi et al. 2014) ground-based infra-red, 
Spitzer mid-infrared, and with deep follow-up spectroscopy 
(e.g. Wang et al. 2015, Karman et al. 2015). Lens models of the Frontier Field clusters
are being produced and constantly updated by several different teams, and can be found at the Hubble 
 Frontier Field website.
\footnote{
https://archive.stsci.edu/prepds/frontier/lensmodels/}. According to Richard
et al. (2014), between 17\% to 38\% of the area in the image plane in each of
the Frontier Fields cluster pointings benefits of a magnification $\mu>10$.
This implies that galaxies 
as faint as $M_{UV}\approx -12.8$, -13.3, -13.6 (assuming $\mu=50$), are detectable at
$z=$3, 4, 5, respectively, over a total of ~7 sq. arcmin.  The Frontier Field will thus soon
allow to extend the investigation of the ultra-faint end of UV luminosity functions at
$z\approx 3-5$ to a $\sim$ 6 times wider area with respect to Alavi et al. (2014), while also
addressing any concern due to cosmic variance thanks to the observation of
six independent pointings, which should reduce it by a factor $\sim 1/\sqrt{6}$. 

Such an observational program will allow to decrease the statistical uncertainty on the measured density of ultra-faint UV galaxies by a factor 3 at least. The effects of such an improvement in the statistics  can be appreciated from inspection of Fig. 3. As an example, assuming  the same central value for the observed number density, rising the lower limit (solid thick line) by a logarithmic factor 0.4 would provide a constraint $m_X\geq 2.3$ keV on the mass of WDM particles. 
\begin{acknowledgements}
We thank the referee for helpful comments that helped to improve the manuscript. 
NM thanks the Observatoire de Paris LERMA - CIAS for the invitation and the kind hospitality.  MC acknowledge the contribution of the FP7 SPACE project ASTRODEEP (Ref.No: 312725), supported by the European Commission. 
This work was partially supported by PRIN INAF 2012.\\ 
\end{acknowledgements}

\newpage

\newpage


\end{document}